\documentclass[11pt,aps, prd, superscriptaddress,nofootinbib]{revtex4-2}
\usepackage[utf8]{inputenc}

\usepackage[T1]{fontenc} 
\usepackage{amsmath}
\usepackage{dsfont}
\usepackage[utf8]{inputenc}
\usepackage{feynmp-auto}
\usepackage{slashed}
\usepackage{bm}
\usepackage{enumerate}

\usepackage{longtable}
\usepackage{multirow}

\usepackage{hhline}
\usepackage{rotating}
\usepackage{array}
\usepackage{float}
\usepackage{afterpage}
\usepackage{anyfontsize}
\usepackage{enumitem}

\usepackage[dvipsnames,table]{xcolor}

\usepackage{arydshln}
\usepackage{blkarray}
\usepackage{physics}
\usepackage{cleveref}
\usepackage{mathtools}
\usepackage{xspace}
\usepackage{setspace}

\usepackage[most]{tcolorbox}
\usepackage{bbold}

  {}

\def\L{{\cal L}}

\def\S{{\cal S}}

\def\eg{\emph{e.g.}}

\allowdisplaybreaks

\newcommand\snowmass{\begin{center}\rule[-0.2in]{\hsize}{0.01in}\\\rule{\hsize}{0.01in}\\
\vskip 0.1in Submitted to the  Proceedings of the US Community Study\\ 
on the Future of Particle Physics (Snowmass 2021)\\ 
\rule{\hsize}{0.01in}\\\rule[+0.2in]{\hsize}{0.01in} \end{center}}

\begin{document}

\snowmass
\vspace{15pt}
\title{Snowmass White Paper:\\Effective Field Theory Matching and Applications}
\author{Timothy Cohen}
\author{Xiaochuan Lu}
\affiliation{Institute for Fundamental Science, University of Oregon, Eugene, OR 97403, USA}
\author{Zhengkang Zhang}
\affiliation{Department of Physics, University of California, Santa Barbara, CA 91106, USA}

\begin{abstract}
    Mapping UV theories onto low energy effective descriptions is a procedure known as matching.  The last decade has seen tremendous progress in the development of new tools for efficiently performing matching calculations, by relying on so-called functional methods. 
    This white paper summarizes the status of functional matching. 
    Specifically, matching for relativistic theories is a fully solved problem up to one-loop order in perturbation theory, and to arbitrary order in the effective field theory expansion. 
    A streamlined prescription that has been partially automated facilitates the application of functional matching to phenomenological studies in the Standard Model EFT framework.
\end{abstract}

\maketitle

\newpage
\section{Executive Summary}

The paradigm of Effective Field Theory (EFT) has a wide range of applications across many areas of physics.  An EFT description emerges when one is analyzing a physical setting that involves a large hierarchy of dimensionful scales.  For example, this could be a theory where there is a hierarchy between the masses of the particles $m \ll M$, in which case the EFT can be derived by ``integrating out'' the heavy state with mass $M$.  The goal of the EFT is to then compute predictions systematically as an expansion in the so-called power counting parameter $m/M$.  See \eg\ Refs.~\cite{Rothstein:2003mp, Skiba:2010xn, Petrov:2016azi, deFlorian:2016spz, Brivio:2017vri, Manohar:2018aog, Neubert:2019mrz, Cohen:2019wxr, Penco:2020kvy} for reviews.

In this white paper, we will be largely focused on the example of interpreting the Standard Model as an EFT (SMEFT), although much of what we discuss is applicable to a variety of other systems.  To write down the Lagrangian for SMEFT, we start with the Standard Model (which only includes operators up to mass dimension four), and add higher dimensional operators built out of the Standard Model fields that are consistent with the Standard Model gauge invariance.  These higher dimensional operators are suppressed by a new physics scale $M$, and the power counting is defined as an expansion in inverse powers of $M$.  Therefore, the operators in SMEFT can be organized in terms of their canonical mass dimension.  It is often said that SMEFT is ``model-independent,'' in that it can capture the influence of a very broad class of low energy signatures that are due to the presence of new heavy degrees of freedom.

As the focus of the LHC discovery program has begun to shift more effort from direct searches for new particles to hunting for indirect signs of beyond the Standard Model physics, the role of SMEFT has become more prominent.  SMEFT parameterizes possible deviations from the Standard Model predictions as an expansion in $1/M$.  So it is reasonable to interpret the measurements being made at the LHC in terms of constraints on the coefficients of the operators that appear in SMEFT (truncated to a fixed order in $1/M$).  Once we have these constraints, the obvious question is how to understand what we learn from them, especially given the large dimensionality of the EFT parameter space.  That is where EFT matching comes in.

Given a UV extension of the Standard Model that includes one or more new heavy states, EFT matching refers to the procedure of computing the coefficients of EFT operators (including threshold corrections) in terms of the UV parameters.  A typical UV model includes far fewer parameters than the EFT.  Therefore, matching facilitates repurposing constraints on the EFT parameters as constraints on the properties of a UV model.  Clearly, a systematic (and ideally automated) approach to matching would dramatically increase the utility of the LHC program.

The theory community has made tremendous progress towards this ambitious goal in the last decade.  As we will summarize in this white paper, a streamlined framework for matching calculations that include effects up to one-loop order and an arbitrary order (even all-order in some special cases) in the power counting has been fully developed, and first steps toward automation have been accomplished.  The key insight has been to reformulate the question using so-called functional methods, where one evaluates the path integral directly~\cite{Gaillard:1985uh,Cheyette:1987qz,Chan:1986jq,Henning:2014gca,Henning:2014wua,Drozd:2015rsp,Henning:2016lyp,Ellis:2016enq,Fuentes-Martin:2016uol,Zhang:2016pja,Cohen:2020fcu}.  

Functional matching at tree-level is the familiar idea of solving for the equations of motion (EOMs) for the heavy fields, plugging this solution back into the UV Lagrangian, and expanding the result in $1/M$.  This not only computes the coefficients of the EFT operators, but it also yields the operators themselves!  This property continues beyond tree-level, and is one of the major benefits of the functional matching approach.
We emphasize several attractive features of functional matching in what follows.
\begin{itemize}
    \item {\it Top-down}. Functional matching is completely top-down; EFT operators are obtained as the end result of the calculation rather than needing to be specified at the beginning.  This saves the significant labor required to work out an EFT operator basis in advance.
    \item {\it Systematic}. Functional methods match the generating functionals of amplitudes, thereby ensuring that all amplitudes are systematically matched between the UV theory and the EFT. This eliminates the need to manually select a specific set of amplitudes to compute.
    \item {\it Economic}. Functional matching extracts just the UV information needed for deriving local operator coefficients in the EFT; there is no need to keep track of IR details.
    \item {\it Efficient}. Functional matching is especially efficient for SMEFT applications due to the universal structure of the calculation at one-loop level.
\end{itemize}

On the practical side, there are now computer programs, \texttt{STrEAM}~\cite{Cohen:2020qvb} and \texttt{SuperTracer}~\cite{Fuentes-Martin:2020udw}, which partially automate one-loop functional matching calculations based on the prescription~\cite{Cohen:2020fcu} summarized in Sec.~\ref{sec:prescription}. 
In the future, it would be useful to further develop these programs and interface them with other (SM)EFT automation tools. 
Also, while the recent technical developments of functional matching have focused on relativistic theories (SMEFT in particular), the techniques may be extended to other cases where the EFT power counting is set by a kinematic restriction, provided there is a clear separation between hard and soft modes. 
This has been studied for the Heavy Quark Effective Theory (HQET)~\cite{Cohen:2019btp}, and should be extended to other ``kinematic'' EFTs in the future, \emph{e.g.}~Soft Collinear Effective Theory.
Finally, it would be exciting to extend functional matching beyond one-loop order.  Developing and applying the formalism of functional matching is clearly a rich subject with many interesting future directions to pursue.

\section{Functional Matching}
\label{sec:prescription}

We now briefly summarize the methodology of EFT matching for relativistic quantum field theories. 
Consider a UV theory, $\L_\text{UV}[\varphi]$, where there is a mass hierarchy among the fields $\varphi$:
\vspace{-2pt}
\begin{equation}
\varphi = \left(\Phi, \phi\right) \,, \qquad\text{with}\qquad m_\Phi\gg m_\phi \,.
\end{equation}
We would like to integrate out the heavy fields $\Phi$ to obtain an EFT for the light fields $\phi$ at low energies, $\L_\text{EFT}[\phi]$, as an expansion in $1/m_\Phi$.

The standard approach to EFT matching is to equate scattering amplitudes in the low energy limit computed (typically via Feynman diagrams) in the UV theory and in the EFT.
In this approach, one first works out a basis of EFT operators and assign them unknown coefficients $\{c_i\}$.
Then, one needs to identify a set of amplitudes that must be computed twice, both using the UV theory and using the EFT, in order to solve for the EFT operator coefficients $\{c_i\}$.
This procedure typically requires significant case-by-case human intervention, and involves conceptually separate tasks such as keeping track of IR details in amplitude calculations.

This standard approach can be implemented systematically and universally by equating the generating functionals of the amplitudes. Concretely, we require that the one-(light-)particle-irreducible (1(L)PI) effective actions be the same in the two theories \cite{Georgi:1991ch, Georgi:1992xg}:
\begin{equation}
\Gamma_\text{EFT}^\text{1PI}[\phi] = \Gamma_\text{UV}^\text{1LPI}[\phi] \,.
\label{eqn:MatchingCondition}
\end{equation}
Since the analytical expression of the 1(L)PI effective action in terms of the Lagrangian is known up to one-loop level, this functional matching condition can be solved analytically to provide us with an explicit expression of the EFT Lagrangian $\L_\text{EFT}[\phi]$ directly in terms of the UV Lagrangian $\L_\text{UV} [\Phi,\phi]$. This allows us to avoid the tedious task of establishing an EFT operator basis in advance, and also saves us the effort in judiciously selecting a set of amplitudes to compute, making the matching calculation dramatically simplified and more systematic.

At tree level, functional matching is achieved be simply substituting in the solution to the classical EOMs for the heavy fields: 
\begin{equation}
\L_\text{EFT}^\text{(tree)}[\phi] = \L_\text{UV} [\Phi,\phi] \big|_{\Phi=\Phi_\text{c}[\phi]} \,, \quad\text{where}\quad \Phi_\text{c}[\phi] \quad\text{solves}\quad
\frac{\delta \S_\text{UV}[\Phi,\phi]}{\delta\Phi} \bigg|_{\Phi=\Phi_c[\phi]} = 0 \,.
\end{equation}
At one-loop level, applying the method of regions~\cite{Beneke:1997zp, Smirnov:2002pj} one finds the following elegant general analytical solution \cite{Henning:2016lyp, Ellis:2016enq, Fuentes-Martin:2016uol, Zhang:2016pja}
\begin{equation}
\int \dd^d x \,\L_\text{EFT}^\text{(1-loop)}[\phi]
\;=\; \frac{i}{2}\, \text{STr}\log \Biggl(-\frac{\delta^2 \S_\text{UV}}{\delta\varphi^2}\biggr|_{\Phi=\Phi_\text{c}[\phi]}\Biggr)\Biggr|_\text{hard} \,.
\label{eq:S_EFT_loop}
\end{equation}
Here the hard region contribution is obtained by expanding the loop integrands assuming the loop momentum $q\sim m_\Phi\gg m_\phi$ before performing the integration using dimensional regularization.
The intuition is that a highly virtual loop with momentum beyond the EFT regime should be encoded by local operators in the EFT.
Because only hard region contributions are kept, this functional approach fully disentangles matching, which deals with UV information, from the IR aspects of amplitude calculations.

For any relativistic UV theory containing generic (non-derivative and derivative) interactions among scalars, fermions, and vector fields, the functional derivative in Eq.~\eqref{eq:S_EFT_loop} can be split into an inverse propagator part $\bm{K}$ and an interaction part $\bm{X}$ (see Ref.~\cite{Cohen:2020fcu} for details):
\begin{equation}
-\frac{\delta^2 \S_\text{UV}}{\delta\varphi^2}\biggr|_{\Phi=\Phi_\text{c}[\phi]} = \bm{K}-\bm{X} \,.
\label{eq:KX}
\end{equation}
This leads to the following general organization of the one-loop matching result
\begin{equation}
\hspace{-8pt}\int \dd^d x \,\L_\text{EFT}^\text{(1-loop)}[\phi] = \frac{i}{2}\,\text{STr}\log \bm{K} \Bigr|_\text{hard}
- \frac{i}{2}\sum_{n=1}^\infty \frac{1}{n} \,\text{STr} \Bigl[\bigl(\bm{K}^{-1}\bm{X}\bigr)^n\Bigr]\Bigr|_\text{hard} \,,
\label{eq:separate_STr}
\end{equation}
where the two terms are respectively termed ``log-type'' and ``power-type'' supertraces. For any given matching calculation, these supertraces can be systematically enumerated graphically~\cite{Cohen:2020fcu}, and then evaluated using the Covariant Derivative Expansion (CDE) technique~\cite{Gaillard:1985uh, Chan:1986jq, Cheyette:1987qz, Henning:2014wua}. This completes the functional matching approach up to one-loop level, for which a pedagogical step-by-step prescription is available in Ref.~\cite{Cohen:2020fcu}.

\section{Application: Universal One-Loop Effective Action}
\label{sec:uolea}

An especially powerful application of functional matching is to derive a so-called Universal One-Loop Effective Action (UOLEA)~\cite{Henning:2014wua, Drozd:2015rsp, Ellis:2016enq, Ellis:2017jns, Summ:2018oko, Kramer:2019fwz, Ellis:2020ivx, Angelescu:2020yzf}. 
The idea is that broad classes of UV theories share common forms of $\bm{K}$ and $\bm{X}$, which allow the functional matching calculation to proceed in a largely UV theory-independent way, resulting in a universal expression for the EFT. 
With this universal expression derived once and for all, we can use it to quickly obtain matching results for specific UV theories by simply substituting in concrete interaction terms.

The UOLEA was first worked out in the minimal case of heavy-only bosonic loops with non-derivative interactions, where the EFT up to dimension six takes a simple form of just 19 terms~\cite{Henning:2014wua, Drozd:2015rsp}. 
Subsequent development of functional matching techniques led to additional UOLEA terms that may arise beyond this minimal case being calculated and tabulated~\cite{Ellis:2016enq, Ellis:2017jns, Summ:2018oko, Kramer:2019fwz, Ellis:2020ivx, Angelescu:2020yzf}.

\section{Application: Matching UV Models onto EFTs}
\label{sec:examples}

The functional approach summarized in Sec.~\ref{sec:prescription} has found numerous applications in matching specific UV models onto SMEFT, with the UOLEA discussed in Sec.~\ref{sec:uolea} further assisting the process in some cases. Examples include integrating out superpartners in the Minimal Supersymmetric Standard Model (MSSM)~\cite{Henning:2014gca,Henning:2014wua,Drozd:2015rsp,Huo:2015nka,Wells:2017vla,Han:2017cfr}, a heavy singlet scalar~\cite{Henning:2014wua,Ellis:2017jns,Jiang:2018pbd,Cohen:2020fcu}, electroweak triplet scalar~\cite{Henning:2014wua,Henning:2016lyp,Ellis:2016enq,Fuentes-Martin:2016uol,Zhang:2016pja}, vectorlike fermions~\cite{Huo:2015exa}, scalar leptoquarks~\cite{Fuentes-Martin:2020udw, Dedes:2021abc}, triplet vector boson~\cite{Brivio:2021alv}, type-I, -II and -III neutrino seesaw models~\cite{Zhang:2021jdf, Du:2022vso, Li:2022ipc}, as well as a survey of 16 different single scalar extensions of the Standard Model~\cite{Anisha:2021hgc}.

There are also example applications of the functional formalism where specific UV theories are matched onto EFTs other than SMEFT, such as HQET~\cite{Cohen:2019btp}, Higgs EFT (result being all-order in the fields)~\cite{Cohen:2020xca}, and also Axion EFT~\cite{Quevillon:2021sfz}.

These results are highly relevant for studying the phenomenology of various new physics models.
For example, precision Higgs coupling measurements are among the most important goals of the LHC and future colliders. 
There are theoretically compelling scenarios ({\it e.g.}\ trans-TeV supersymmetry that realizes $b$-$\tau$ Yukawa unification~\cite{Wells:2017vla}) where Higgs coupling deviations will be the first sign of new physics beyond the Standard Model. 
In such cases, EFT matching calculations provide the bridge between EFT operators measured in the IR and new physics parameters in the UV, which then offers valuable guidance for direct searches.

\section{Automated Tools}
\label{sec:tools}

In light of the functional formalism summarized in Sec.~\ref{sec:prescription}, various automated tools have been developed in the past few years to facilitate EFT matching calculations.
Supertrace evaluation is arguably the most technically involved step in the functional approach to EFT matching. The Mathematica packages \texttt{STrEAM}~\cite{Cohen:2020qvb} and \texttt{SuperTracer}~\cite{Fuentes-Martin:2020udw} are designed to handle this task. Automating the CDE technique, these packages evaluate any given supertrace that one could possibly encounter in an EFT matching calculation, and outputs the effective operators (with their coefficients) up to any specified mass dimension truncation.
\texttt{SuperTracer} also automatically enumerates the relevant supertraces in a given EFT matching calculation; it is designed to be part of a fully automated master program of EFT matching termed \texttt{Matchete} (see Ref.~\cite{Fuentes-Martin:2020udw} for its flow chart).
Programs that partially compute one-loop EFT Lagrangians by coding up the UOLEA results up to mass dimension six also exist; see \texttt{CoDEx}~\cite{DasBakshi:2018vni}.
Beyond Mathematica packages, a Python library \texttt{MatchingTools}~\cite{Criado:2017khh} is also available for addressing EFT matching at tree level using functional approach.
Finally, plain Feynman diagram approach to EFT matching has also been automated by the package \texttt{Matchmakereft}~\cite{Carmona:2021xtq}.

\vspace{20pt}
{\bf Acknowledgments.}
T.C.\ and X.L.\ are supported by the U.S.\ Department of Energy, under grant number DE-SC0011640. 
Z.Z.\ is supported by the U.S.\ Department of Energy under the grant DE-SC0011702.

\addcontentsline{toc}{section}{\protect\numberline{}References}%
\bibliographystyle{utphys}
\bibliography{ref}

\end{document}